\documentclass{article}
\usepackage{amsmath}
\usepackage{amsfonts}
\usepackage{amssymb}
\usepackage{graphicx}
\usepackage{dsfont}
\usepackage[lined,boxed,commentsnumbered]{algorithm2e}

\providecommand{\U}[1]{\protect\rule{.1in}{.1in}}
\providecommand{\U}[1]{\protect\rule{.1in}{.1in}}
\newtheorem{theorem}{Theorem}

\newtheorem{remark}[theorem]{Remark}

\begin{document}

\title{Conditional inference in parametric models}
\author{Michel Broniatowski$^{(1)}$, Virgile Caron$^{(1)}$ \\
$^{(1)}$ LSTA, Universit\'{e} Pierre et Marie Curie, Paris, France}
\maketitle

\begin{abstract}
This paper presents a new approach to conditional inference, based on the
simulation of samples conditioned by a statistics of the data. Also an
explicit expression for the approximation of the conditional likelihood of
long runs of the sample given the observed statistics is provided. It is
shown that when the conditioning statistics is sufficient for a given
parameter, the approximating density is still invariant with respect to the
parameter. A new Rao-Blackwellisation procedure is proposed and simulation
shows that Lehmann Scheff\'{e} Theorem is valid for this approximation.
Conditional inference for exponential families with nuisance parameter is
also studied, leading to Monte carlo tests. Finally the estimation of the
parameter of interest through conditional likelihood is considered.
Comparison with the parametric bootstrap method is discussed.

\end{abstract}

\section{Introduction and context}

This paper explores conditional inference in parametric models. A
comprehensive overview on this area is the illuminating review paper by Reid
(1995) \cite{Reid1995}. Our starting point is as follows: given a model $%
\mathcal{P}$ defined as a collection of continuous distributions $P_{\theta} 
$ on $\mathbb{R}^{d}$ , with density $p_{\theta}$ where the parameter $%
\theta $ belongs to some subset $\Theta$ in $\mathbb{R}^{s}$ and given a
sample of independent copies of a random variable with distribution $%
P_{\theta_{T}}$ for some unknown value $\theta_{T}$ of the parameter, we
intend to provide some inference about $\theta_{T}$ conditioning on some
statistics of the data. The situations which we have in mind are of two
different kinds.

The first one is the Rao-Blackwellisation of estimators, which amounts to
reduce the variance of an unbiased estimator by conditioning on any
statistics; it is a known fact that such method reduces its variance; when
the conditioning statistics is complete and sufficient for the parameter
then this procedure provides optimal reduction, as stated by Lehmann-Scheff%
\'{e} Theorem. This realm of questions is the motivation for the first part
of this paper:

\begin{enumerate}
\item is it possible to provide good approximations for the density of a
sample conditioned on a given statistics, and, when applied for a model
where some sufficient statistics for the parameter is known, does
sufficiency w.r.t. the parameter still holds for the approximating density?

\item in the case when the first question has positive answer, is it
possible to simulate samples according to the approximating density, and to
propose some Rao-Blackwellised version for a given preliminary estimator?
Also we would hope that the proposed method would be feasible, that the
programming burden would be light, that the run time for this method be
short, and that the involved techniques would keep in the range of globally
known ones by the community of statisticians.
\end{enumerate}

The second application of conditional inference pertains to the role of
conditioning in models with nuisance parameters. There is a huge
bibliography on this topic, some of which will be considered in details in
the sequel. The usual frame for this field of problems is the exponential
families one, for reasons related both with the importance of these models
in applications and on the role of the concept of sufficiency when dealing
with the notion of nuisance parameter. Conditioning on a sufficient
statistics for the nuisance parameter produces a new exponential family,
which gets free of this parameter, and allows for simple inference on the
parameter of interest, at least in simple cases. This will also be
discussed, since the reality, as known, is not that simple, and since so
many complementary approaches have been developped over decades in this
area. Using the approximation of the conditional density in this context and
performing simulations yields Monte Carlo tests for the parameter of
interest, free from the nuisance parameter. Also conditional maximum
likelihood estimators will be produced. Comparison with the parametric
bootstrap will also be discussed.

\bigskip

This paper is organized as follows. Section 2 describes a general
approximation scheme for the conditional density of long runs of subsamples
conditioned on a statistics, with explicit formulas. The (rather lengthy)
proof of the main result of this section is presented in Broniatowski and
Caron(2011)\cite{BroniaCaronAAP2011}. Discussion about implementation is
provided. Section 3 presents two aspects of the approximating conditional
scheme: we first show on examples that sufficiency is kept under the
approximating scheme and, second, that this yields to an easy
Rao-Blackwellisation procedure. An illustration of Lehmann-Scheff\'{e}
Theorem is presented. Section 4 deals with models with nuisance parameters
in the context of exponential families. We have found it useful to spend a
few paragraphs on bibliographical issues. We address Monte Carlo tests based
on the simulation scheme; in simple cases its performance is similar to that
of parametric bootstrap; however conditional simulation based tests improve
clearly over parametric bootstrap procedure when the test pertains to models
for which the likelihood is multimodal with respect to the nuisance
parameter; an example is provided. Finally we consider conditioned maximum
likelihood based on the approximation of the conditional density; in simple
cases its performance is similar to that of unconditional likelihood;
however when the preliminary estimator of the nuisance is difficult to
obtain, for example when it depends strongly on some initial point for a
Newton-Raphson routine (this is indeed a very common situation), then, by
the very nature of sufficiency, conditional inference based on the proxy of
the conditional likelihood performs better; this is illustrated with
examples.

\section{The approximate conditional density of the sample}

Most attempts which have been proposed for the approximation of conditional
densities stem from arguments developped in Lehmann (1986)\cite{Lehmann1986}
for inference on the parameter of interest in models with nuisance
parameter; however the proposals in this direction hinge at the
approximation of the distribution of the sufficient statistics for the
parameter of interest given the observed value of the sufficient statistics
of the nuisance parameter. We will present some of these proposals in the
section devoted to exponential families. To our knowledge, no attempt has
been made to approximate the conditional distribution of a sample (or of a
long subsample) given some observed statistics.

However, generating samples from the conditional distribution itself (such
samples are often called co-sufficient samples, following Lockhart et
al.(2007) \cite{LockhartOreillyStephens2007}) has been considered by many
authors; see for example Engen and Lillegard (1997)\cite{EngenLillegard1997}%
, Lindqvist et al. (2003)\cite{Lindqvist2003} and references therein, and
Lindqvist and Taraldsen (2005)\cite{LinqvistTaraldsen2005}.

In Engen and Lillegard (1997)\cite{EngenLillegard1997}, simulating
exponential or normal samples under the given value of the empirical mean is
proposed. For example under the exponential distribution $Exp(\theta),$ the
minimal sufficient statistics for $\theta$ is the sum of the observations,
say $t_{n};$ a co-sufficient sample $x^{\ast}$ can be created by generating
an $x^{^{\prime}}$-sample from $Exp(1)$ and taking $x_{i}^{\ast}=x_{i}^{^{%
\prime }}t_{n}/\overline{x}^{^{\prime}}.$ However, this approach may be at
odd in simple cases, as for the Gamma density in the non exponential case.

Lockhart et al. (2007)\cite{LockhartOreillyStephens2007} proposed a
different framework based on the Gibbs sampler, simulating the conditioned
sample one at a time through a sequential procedure. The example which is
presented is for the Gamma distribution under the empirical mean, but it
seems to perform well, for location parameter, when the true parameter is in
some range, therefore not uniformly on the model. Their paper contains a
comparative study with the parametric bootstrap procedure (introduced by
Efron (1979)\cite{efron1979}) for similar problems. In a simple case, they
argue favorably for both methods. We will turn back to parametric bootstrap
in relation with conditional likelihood estimators, in the last section of
this paper.

Other techniques have been developped in specific cases: for the inverse
gaussian distribution see O'Reilly and Gravia-Medrano (2006)\cite%
{OreillyGravia2006}, Cheng (1984) \cite{chen1984}; for the Weibull
distribution see Lockhart et Stephens (1994)\cite{LockartStephens1994}. No
unified technique exists in the litterature which would work under general
models.

\subsection{Approximation of conditional densities}

\subsubsection{Notation and hypotheses}

For sake of clearness we consider the case when the model $\mathcal{P}$ is a
family of distributions on $\mathbb{R}$. Extension to $\mathbb{R}^{d},d>1$
can be achieved in the same way, using similar results developed in futur
work.

Denote $\mathbf{X}_{1}^{n}:=\left( \mathbf{X}_{1}\mathbf{,..,X}_{n}\right)$
a set of $n$ independent copies of a real random variable $\mathbf{X}$ with
density $p_{\mathbf{X},\theta_{T}}$ on $\mathbb{R}$. Let $\mathbf{x}%
_{1}^{n}:=\left(\mathbf{x}_{1},...,\mathbf{x}_{n}\right)$ denote the
observed values of the data, each $\mathbf{x}_{i}$ resulting from the
sampling of $\mathbf{X}_{i}.$ Define the r.v. $\mathbf{U}:=u\left(\mathbf{X}%
\right)$ and $\mathbf{U}_{1,n}:=u\left( \mathbf{X}_{1}\right) +...+u\left( 
\mathbf{X}_{n}\right)$ where $u$ is a real-valued measurable function on $%
\mathbb{R}$, and, accordingly, $u_{1,n}:=u\left( \mathbf{x}_{1}\right)
+...+u\left( \mathbf{x}_{n}\right).$ Denote $p_{\mathbf{U},\theta_{T}}$ the
density of the r.v. $\mathbf{U}.$ We consider approximations of the density
of the vector $\mathbf{X}_{1}^{k}=\left(\mathbf{X}_{1}\mathbf{,..,X}%
_{k}\right) $ on $\mathbb{R}$ when $\mathbf{U}_{1,n}=u_{1,n}.$ It will be
assumed that the observed value $u_{1,n}$ is "typical", in the sense that it
keeps in the range of the iterated logarithm law order of magnitude (for
large $n$). Large deviation cases could also be handled, but conditional
inference is based on the implicit assumption that such cases are excluded
from the analysis. We hence assume 
\begin{equation}  \label{LIL}
\lim\sup_{n\rightarrow\infty}\frac{\left\vert u_{1,n}-nE[u\left( \mathbf{X}%
\right) ]\right\vert }{\sqrt{Var(u\left( \mathbf{X}\right) )}\sqrt{%
2n\log\log n}}=1.  \tag{LIL}
\end{equation}

We propose an approximation for 
\begin{equation}
p_{u_{1,n},\theta_{T}}\left( x_{1}^{k}\right) :=p_{\theta_{T}}(x_{1}^{k}|%
\mathbf{U}_{1,n}=u_{1,n})
\end{equation}
where $\mathbf{X}_{1}^{k}:=\left( \mathbf{X}_{1}\mathbf{,..,X}_{k}\right) $
and $k:=k_{n}$ is an integer sequence such that 
\begin{equation}  \label{K1}
0\leq \lim \sup_{n\rightarrow \infty }k/n\leq {1}  \tag{K1}
\end{equation}
together with 
\begin{equation}  \label{K2}
\lim_{n\rightarrow\infty}n-k=\infty  \tag{K2}
\end{equation}
which is to say that we approximate $p_{u_{1,n},\theta_{T}}\left(x_{1}^{k}%
\right)$ on long runs. The rule which define the value of $k$ for a given
accuracy of the approximation is stated in section 3.2 of Broniatowski and
Caron(2011) \cite{BroniaCaronAAP2011}. \newline
The hypotheses pertaining to the function $u$ and the r.v. $\mathbf{U}%
=u\left(\mathbf{X}\right) $ are as follows

\begin{enumerate}
\item $u$ is real valued and the characteristic function of the random
variable $\mathbf{U}$ is assumed to belong to $L^{r}$ for some $r\geq{1}.$

\item The r.v. $\mathbf{U}$ is supposed to fulfill the Cramer condition: its
moment generating function satisfies 
\begin{equation*}
\phi_{\mathbf{U}}(t):=E\exp t\mathbf{U}<\infty
\end{equation*}
for $t$ in a non void neighborhood of $0.$
\end{enumerate}

Define the functions $m(t),s^{2}(t)$ and $\mu_{3}(t)$ as the first, second
and third derivatives of $\log\phi_{\mathbf{U}}(t).$ Denote 
\begin{equation*}
\pi_{u,\theta_{T}}^{\alpha}(x):=(x):=\frac{\exp tu(x)}{\phi _{\mathbf{U}}(t)}%
p_{\mathbf{X},\theta_{T}}\left( x\right)
\end{equation*}
with $m(t)=\alpha$ and $\alpha$ belongs to the support of $P_{\mathbf{X}%
,\theta_{T}}$, the distribution of $\mathbf{X}$. The density $%
\pi_{u,\theta_{T}}^{\alpha}$ is the \textit{tilted} density with parameter $%
\alpha.$ Also it is assumed that this latest definition of $t$ makes sense
for all $\alpha$ in the support of $\mathbf{X}.$ Conditions on $\phi_{%
\mathbf{U}}(t)$ which ensure this fact are referred to as \textit{steepness
properties}, and are exposed in Barndorff-Nielsen(1978)\cite%
{BarndorffNielsen1978}, p153.

We introduce a positive sequence $\epsilon_{n}$ which satisfies

\begin{align}
\lim_{n\rightarrow\infty}\epsilon_{n}\sqrt{n-k}=\infty  \tag{E1}  \label{EE1}
\\
\lim_{n\rightarrow\infty}\epsilon_{n}\left(\log n\right)^{2}=0.  \tag{E2}
\label{EE2}
\end{align}

\subsection{The proxy of the conditional density of the sample}

We recusively define the density $g_{u_{1,n},\theta_{T}}(x_{1}^{k})$ on $%
\mathbb{R}^{k}$, which approximates $p_{u_{1,n},\theta_{T}}\left(x_{1}^{k}%
\right)$ sharply with relative error smaller than $\epsilon_{n}\left(\log
n\right)^{2}.$ The subsript $\theta_{T}$ will be omitted when there is no
ambiguity about the value of the parameter.

Set 
\begin{equation*}
m_{0}:=u_{1,n}/n.
\end{equation*}%
and 
\begin{equation*}
g_{0}(\left. x_{1}\right\vert x_{0}):=\pi _{u}^{m_{0}}(x_{1})
\end{equation*}%
with $x_{0}$ arbitrary, and for $1\leq i\leq k-1$ define the density $%
g(\left. x_{i+1}\right\vert x_{1}^{i})$ recursively as follows.

Set $t_{i}$ the unique solution of the equation 
\begin{equation}
m_{i}:=m(t_{i})=\frac{u_{1,n}-u_{1,i}}{n-i}  \label{mi}
\end{equation}%
where $u_{1,i}:=u(x_{1})+...+u(x_{i}).$ The tilted adaptive family of
densities $\pi _{\mathbf{X}}^{m_{i}}$ is the basic ingredient of the
derivation of approximating scheme. Let 
\begin{equation*}
s_{i}^{2}:=\frac{d^{2}}{dt^{2}}\left( \log E_{\pi _{u}^{m_{i}}}\exp tu\left( 
\mathbf{X}\right) \right) \left( 0\right) 
\end{equation*}%
and 
\begin{equation*}
\mu _{j}^{i}:=\frac{d^{j}}{dt^{j}}\left( \log E_{\pi _{u}^{m_{i}}}\exp
tu\left( \mathbf{X}\right) \right) \left( 0\right) ,\text{ }j=3,4
\end{equation*}%
which are the second, third and fourth cumulants of $\pi ^{m_{i}}.$ Let 
\begin{equation}
g(\left. x_{i+1}\right\vert x_{1}^{i})=C_{i}p_{\mathbf{X,\theta }%
_{T}}(x_{i+1})\mathfrak{n}\left( \alpha \beta ,\beta ,u\left( x_{i+1}\right)
\right)   \label{g-i}
\end{equation}%
where $\mathfrak{n}\left( \mu ,\tau ,x\right) $ is the normal density with
mean $\mu $ and variance $\tau $ at $x$. Here 
\begin{equation}
\beta =s_{i}^{2}\left( n-i-1\right)   \label{a}
\end{equation}%
\begin{equation}
\alpha =t_{i}+\frac{\mu _{3}^{i}}{2s_{i}^{4}\left( n-i-1\right) }  \label{b}
\end{equation}%
and the $C_{i}$ is a normalizing constant.

Define 
\begin{equation}
g_{u_{1,n}}(x_{1}^{k}):=g_{0}(\left. x_{1}\right\vert
x_{0})\prod_{i=1}^{k-1}g(\left. x_{i+1}\right\vert x_{1}^{i}).  \label{g_s}
\end{equation}
It holds

\begin{theorem}
\label{Thm:egal_stat} Assume (\ref{K1},\ref{K2}) together with (\ref{EE1},%
\ref{EE2}). Then (i) 
\begin{equation*}
p_{u_{1,n}}(x_{1}^{k})=g_{u_{1,n}}(x_{1}^{k})(1+o_{P_{u_{1,n}}}(\epsilon
_{n}\left( \log n\right) ^{2}))
\end{equation*}

and (ii) 
\begin{equation*}
p_{u_{1,n}}(x_{1}^{k})=g_{u_{1,n}}(x_{1}^{k})(1+o_{G_{u_{1,n}}}(\epsilon
_{n}\left( \log n\right) ^{2})).
\end{equation*}
\end{theorem}

For the proof, see Broniatowski and Caron (2011) \cite{BroniaCaronAAP2011}.

Statement (i) means that the conditional likelihood of any long sample path $%
\mathbf{X}_{1}^{k}$ given $\mathbf{U}_{1,n}=u_{1,n}$ can be approximated by $%
g_{u_{1,n}}(\mathbf{X}_{1}^{k})$ with a small relative error on typical
realizations of $\mathbf{X}_{1}^{n}.$

The second statement states that simulating $\mathbf{X}_{1}^{k}$ under $%
g_{u_{1,n}}$ produces runs which could have been sampled under the
conditional density $p_{u_{1,n}}$ since $g_{u_{1,n}}$ and $p_{u_{1,n}}$
coincide sharply on larger and larger subsets of $\mathbb{R}^{k}$ as $n$
increases.

\begin{remark}
Theorem \ref{Thm:egal_stat} states that the density $g_{u_{1,n},\left(
\theta _{T},\eta _{T}\right) }$ on $\mathbb{R}^{k}$ approximates $%
p_{u_{1,n},\left( \theta _{T},\eta _{T}\right) }$ on the sample $\mathbf{x}%
_{1}^{n}$ generated under $\left( \theta _{T},\eta _{T}\right) .$ However,
in some cases, the r.v.'s $\mathbf{x}_{i}$'s in Theorem \ref{Thm:egal_stat}
may at time be generated under some other parameters, say $\left( \theta
_{0},\eta _{0}\right) .$ Indeed, for direct applications developped in this
paper, Theorem \ref{Thm:egal_stat} have to hold when the sample is generated
under an other sampling scheme. Broniatowski and Caron (2011) \cite%
{BroniaCaronAAP2011} state in Theorem 11 that the approximation scheme holds
true in this case.

Let $\mathbf{Y}_{1}^{n}$ be i.i.d. copies of $\mathbf{Z}$ with distribution $%
Q$ and density $q;$ assume that $Q$ satisfies the Cramer condition $\int
\left( \exp tx\right) q(x)dx<\infty $ for $t$ in a non void neighborhood of $%
0.$ Let $\mathbf{V}_{1,n}:=u\left( \mathbf{Y}_{1}\right) +...+u\left( 
\mathbf{Y}_{n}\right) $ and define 
\begin{equation*}
q_{u_{1,n}}\left( y_{1}^{k}\right) :=q\left( \left. \mathbf{Y}%
_{1}^{k}=y_{1}^{k}\right\vert \mathbf{V}_{1,n}=u_{1,n}\right) 
\end{equation*}%
with distribution $Q_{u_{1,n}}.$ It then holds
\end{remark}

\begin{theorem}
\label{ThmApproxSousAutreTheta} Then, with the same hypotheses and notation
as in Theorem \ref{Thm:egal_stat}, 
\begin{equation*}
p\left( \mathbf{X}_{1}^{k}=Y_{1}^{k}\left\vert \mathbf{U}_{1,n}=u_{1,n}%
\right. \right) =g_{u_{1,n}}(Y_{1}^{k})(1+o_{Q_{u_{1,n}}}(\epsilon
_{n}\left( \log n\right) ^{2})).
\end{equation*}

Also the total variation distance between $Q_{u_{1,n}}$ and $P_{u_{1,n}}$
goes to $0$ as $n$ tends to infinity.
\end{theorem}

\subsection{Comments on implementation}

The simulation of a sample $X_{1}^{k}$ with density $g_{u_{1,n}}$ is fast as
easy. Indeed the r.v. $X_{i+1}$ with density $g\left(
x_{i+1}|x_{1}^{i}\right) $ is obtained through a standard acceptance
-rejection algorithm. When $\theta _{T}$ is unknown, a preliminary estimator
may be used. When $\mathbf{U}_{1,n}$ is sufficient for $p_{u_{1,n}}$ it is
nearly sufficient for its proxy $g_{u_{1,n}}$ (see next section); indeed
changing the value of this preliminary estimator does not alter the
likelihood of the sample; as shown in the simulations developped here after,
any value of $\theta $ can be used; call $\theta ^{\ast }$ the $\theta $
chosen as initial value , using henceforth $p_{\mathbf{X,\theta }^{\ast }}$
instead of $p_{\mathbf{X,\theta }_{T}}$ in (\ref{g-i}). In exponential
families the values of the parameters which appear in the gaussian component
of $g\left( x_{i+1}|x_{1}^{i}\right) $ in (\ref{g-i}) are easily calculated;
note also that due to (\ref{LIL}) the parameters in $\mathfrak{n}\left(
\alpha \beta ,\beta ,u\left( x_{i+1}\right) \right) $ are such that the
dominating density can be chosen for all $i$ as $p_{\mathbf{X,\theta }^{\ast
}}.$ The constant in the acceptance rejection algorithm is then $1/\sqrt{%
2\pi \alpha }.$ This is in contrast with the case when the conditioning
value is in the range of a large deviation with respect to $p_{\mathbf{%
X,\theta }_{T}}$; in this case, which appears in a natural way in Importance
sampling estimation for rare event probabilities, the simulation algorithm
is more complex ; see \cite{BroniaCaronRareevents2011}.

\section{Sufficient statistics and approximated conditional density}

\subsection{Keeping sufficiency under the proxy density}

\label{sec:keep_suff}

The density $g_{u_{1,n}}(y_{1}^{k})$ is used in order to handle Rao
-Blackellisation of estimators or statistical inference for models with
nuisance parameters. The basic property is sufficiency with respect to the
envolved parameter. We show on some examples that $g_{u_{1,n}}(y_{1}^{k})$
defined in (\ref{g_s}) inherits of the invariance with respect to a
parameter when conditioning on a sufficient statistics pertaining to this
parameter.

Consider the Gamma density 
\begin{equation}
f_{\rho,\theta}(x):=\frac{\theta^{-\rho}}{\Gamma(\rho)}x^{\rho-1}\exp
-x/\theta\text{ \ \ \ for }x>0.  \label{gamma}
\end{equation}
As $r$ varies in $\mathbb{R}^{+}$ and $\theta$ is positive, the density runs
in an exponential family $\gamma_{r,\theta}$ with parameters $r:=\rho-1$ and 
$\theta,$ and sufficient statistics $t(x):=\log x$ and $u(x):=x$
respectively for $r$ and $\theta.$ Given \ an i.i.d. sample $%
X_{1}^{n}:=\left( X_{1},...,X_{n}\right) $ with density $f_{r_{T},%
\theta_{T}} $ the resulting sufficient statistics are respectively $%
T_{1,n}:=\log X_{1}+...+\log X_{n}$ and $U_{1,n}:=X_{1}+...+X_{n}.$ We
consider two parametic models $\left(\gamma_{r_{T},\theta},\theta\geq{0}%
\right) $ and $\left(\gamma_{r,\theta_{T}\text{ }},r>0\right) $ respectively
assuming $r_{T}$ or $\theta_{T}$ known.

We first consider sufficiency of $U_{1,n}$ in the first model. The density $%
g_{u_{1,n}}(y_{1}^{k})$ should be free of the current value of the true
parameter $\theta _{T}$ of the parameter under which the data are drawn.
However as appears in (\ref{g_s}) the unknown value $\theta _{T}$ should be
used in its very definition. We show by simulation that whatever the value
of $\theta $ inserted in place of $\theta _{T}$ in (\ref{g_s}) the
likelihood of $X_{1}^{k}$ under $g_{u_{1,n}}$ does not depend upon $\theta .$
We thus observe that $U_{1,n}$ is sufficient for $\theta _{T}$ in the
conditional density approximating $p_{u_{1,n}}$ as should hold as a
consequence of Theorem \ref{Thm:egal_stat} .

Similarly the same fact occurs replacing $\theta_{T}$ by $r_{T}$ in the
model $\left(\gamma_{r,\theta_{T}},r>0\right) .$

\begin{figure}[tbp]
\centering \includegraphics*[scale=0.4]{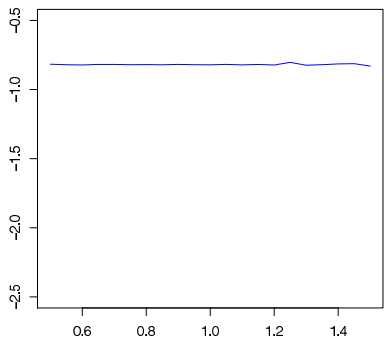}
\caption{Proxy of the conditional likelihood of $X_{1}^{k}$ under $%
g_{U_{1,n}}$ as a function of $r$ for $n=100$ and $k=80$ in the gamma case.}
\label{Figure:Gamma_sufficiency_2}
\end{figure}

In both cases whatever the value of the parameter $\theta$ (Figure \ref%
{Figure:Gamma_sufficiency_1}) or $r$ (Figure \ref{Figure:Gamma_sufficiency_2}%
), the likelihood of $X_{1}^{k}$ remains constant.

\begin{figure}[tbp]
\centering \includegraphics*[ scale=0.4]{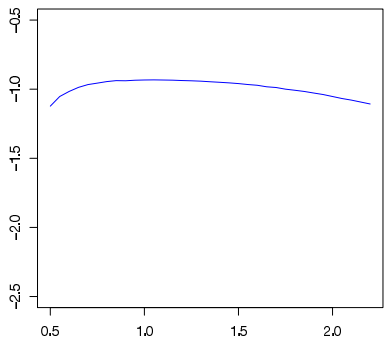}
\caption{Proxy of the conditional likelihood of $X_{1}^{k}$ under $%
g_{T_{1,n}}$ as a function of $\protect\theta$ for $n=100$ and $k=80$ in the
gamma case.}
\label{Figure:Gamma_sufficiency_1}
\end{figure}

\begin{figure}[tbp]
\centering \includegraphics*[ scale=0.4]{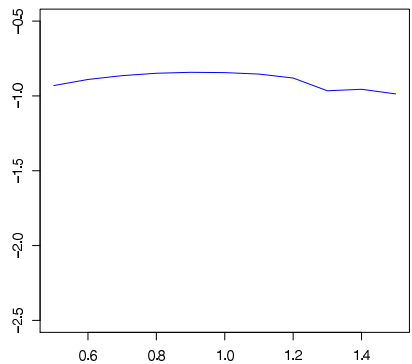}
\caption{Conditional likelihood of $X_{1}^{k}$ under $g_{U_{1,n}}$ as a
function of $\protect\lambda$ for $n=100$ and $k=80$ in the Inverse Gaussian
case.}
\label{Figure:Inverse_Gaussian_sufficiency_2}
\end{figure}

We also consider the Inverse Gaussian distribution with density

\begin{eqnarray}  \label{inverse_gaussienne}
f_{\lambda,\mu}(x):=\left[ \frac{\lambda}{2\pi}\right] ^{1/2}\exp -\frac{%
\lambda\left( x-\mu\right) ^{2}}{2\mu^{2}x} \text{ \ \ \ for }x>0
\end{eqnarray}
with both parameters $\lambda$ and $\mu$ be positive. Given an i.i.d. sample 
$X_{1}^{n}:=\left( X_{1},...,X_{n}\right)$ with density $f_{\mu,\lambda}$,
the resulting sufficient statistics are respectively $%
T_{1,n}:=X_{1}+...+X_{n}$ and $U_{1,n}:=X_{1}^{-1}+...+X_{n}^{-1}.$
Similarly as for the Gamma case we draw the likelihood of a subsample $%
X_{1}^{k}$ under $g_{u_{1,n}}$ with $T_{1,n}:=X_{1}+...+X_{n}$ ,which is a
sufficient statistics for $\mu$ (Figure \ref%
{Figure:Inverse_Gaussian_sufficiency_1}), and upon $%
U_{1,n}:=X_{1}^{-1}+...+X_{n}^{-1}$ which is sufficient for $\lambda$
(Figure \ref{Figure:Inverse_Gaussian_sufficiency_2}). In either cases the
other coefficient is kept fixed at the true value of the parameter
generating the sample. As for the Gamma case these curves show the
invariance of the proxy of the conditional density with respect to the
parameter for which the chosen statistics is sufficient.

\begin{figure}[tbp]
\centering \includegraphics*[ scale=0.4]{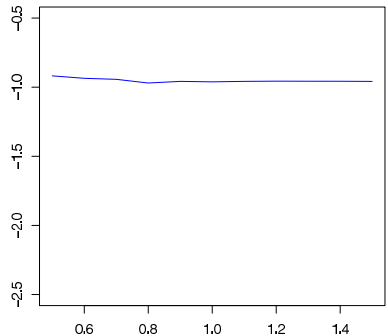}
\caption{Conditional likelihood of $X_{1}^{k}$ under $g_{T_{1,n}}$ as a
function of $\protect\mu$ for $n=100$ and $k=80$ in the Inverse Gaussian
case.}
\label{Figure:Inverse_Gaussian_sufficiency_1}
\end{figure}

\subsection{Rao-Blackwellisation}

Rao-Blackwell Theorem holds regardless of whether biased or unbiased
estimators are used, since it reduces the MSE. Although its statement is
rather weak, in practice, however, the improvement is often enormous. New
interest in Rao-Blackwellisation procedures have risen in the recent years,
conditioning on ancillary variables (see Fraser(2004) \cite{Fraser12004} for
a survey on ancillaries in conditional inference); specific
Rao-Blackwellisation schemes have been proposed by Casella and Robert \cite%
{CasellaRobert1996}, \cite{CasellaRobert1998}, Perron(1999)\cite{Perron1999}%
, Douc and Robert (2010)\cite{robertDouc2011} and Iacobucci et all.(2010) 
\cite{Jacobbucci}. The purpose is to improve the variance of a given
statistics (for example a tail probability) under a \textit{known}
distribution, through a simulation scheme under this distribution; the
ancillary variables used in the simulation process itself are used as
conditioning ones for the Rao-Blackwellisation of the statistics. The
present approach is more classical in this respect, since we do not assume
that the parent distribution is known; conditioning on a sufficient
statistics $\mathbf{U}_{1,n}$ with respect to the parameter $\theta $ and
simulating samples according to the approximating density $g_{u_{1,n}}$ will
produce the improved estimator.

Since $U_{1,n}$ is sufficient for the parameter $\theta$ in $g_{u_{1,n}}$ it
can be used in order to obtain improved estimators of $\theta_{T}$ through
Rao Blackwellization. We shortly illustrate the procedure and its results on
some toy cases. Consider again the Gamma family defined here-above with
canonical parameters $r$ and $\theta$.

First the parameter to be estimated is $\theta_{T}.$ A first unbiaised
estimator is chosen as 
\begin{equation*}
\widehat{\theta}_{2}:=\frac{X_{1}+X_{2}}{2r_{T}}.
\end{equation*}
Given an i.i.d. sample $X_{1}^{n}$ with density $\gamma_{r_{T},\theta_{T}}$
the Rao-Blackwellised estimator of $\widehat{\theta}$ is defined through%
\begin{equation*}
\theta_{RB,2}:=E\left( \left. \widehat{\theta_{2}}\right\vert U_{1,n}\right)
\end{equation*}
whose variance is less than $Var\widehat{\theta_{2}}.$

Consider $k=2$ in $g_{U_{1,n}}(y_{1}^{k})$ and let $\left(
Y_{1},Y_{2}\right) $ be distributed according to $g_{u_{1,n}}(y_{1}^{2}).$
Replications of $\left( Y_{1},Y_{2}\right) $ induce an estimator of $%
\theta_{RB,2}$ for fixed $u_{1,n}.$ Iterating on the simulation of the runs $%
X_{1}^{n}$ produces, for $n=100$ an i.i.d. sample of $\theta_{RB,2}$'s and
the $Var\theta_{RB,2}$ is estimated. The resulting variance shows a net
improvement with respect to the estimated variance of $\widehat{\theta}_{2}.$
It is of some interest to confront this gain in variance as the number of
terms involved in $\widehat{\theta}_{k}$ increases together with $k.$ As $k$
approaches $n$ the variance of $\widehat{\theta}_{k}$ approaches the Cramer
Rao bound. The graph below shows the decay in variance of $\widehat{\theta }%
_{k}.$ We note that whatever the value of $k$ the estimated value of the
variance of $\theta_{RB,k}$ is constant. This is indeed an illustration of
Lehmann-Scheff\'{e}'s theorem.

\begin{figure}[tbp]
\centering \includegraphics*[ scale=0.45]{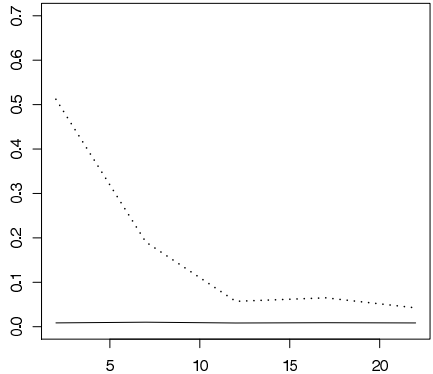}
\caption{Variance of $\widehat{\protect\theta}_{k}$, the initial estimator
(dotted line), along with the variance of $\protect\theta_{RB,k}$, the
Rao-Blackwellised estimator (solid line) with $n=100$ as a function ok $k.$}
\end{figure}

\begin{remark}
Lockhart and O'Reilly (2005) \cite{LOckhart2005} establish, under certain
conditions and for \textit{fixed} $k,$ the asymptotic equivalence of the
plug-in estimate for the distribution $P_{\theta_{ML}}\left(\mathbf{X}%
_{1}^{k}\in B\right) $ and the Rao-Blackwell estimate $P\left(\left. \mathbf{%
X}_{1}^{k}\in B\right\vert \mathbf{U}_{1,n}\right) $ where $\theta_{ML}$ is
the maximum likelihood estimator of $\theta_{T}$ based on the whole sample $%
\mathbf{X}_{1}^{n}$ (this result is known as Moore's conjecture (see
Moore(1973)\cite{Moore1973})). They also provide rates for this convergence.
\end{remark}

\section{Exponential models with nuisance parameters}

\subsection{Conditional inference in exponential families}

We consider the case when the parameter consists in two distinct
subparameters, one of interest denoted $\theta $ and a nuisance component
denoted $\eta $. As is well known, conditioning on a sufficient statistics
for the nuisance parameter produces a new exponential family which is free
of it. Assuming the observed dataset $\mathbf{x}_{1}^{n}:=\left( \mathbf{x}%
_{1},...,\mathbf{x}_{n}\right) $ resulting from sampling of a vector $%
\mathbf{X}_{1}^{n}:=\left( \mathbf{X}_{1},...,\mathbf{X}_{n}\right) $ of
i.i.d. random variables with distribution in the initial exponential model,
and denoting $u\left( \mathbf{x}_{1}^{n}\right) $ a sufficient statistics
for $\eta ,$ simulation of samples under the conditional distribution of $%
\mathbf{X}_{1}^{n}$ given $u\left( \mathbf{X}_{1}^{n}\right) =$ $u\left( 
\mathbf{x}_{1}^{n}\right) $ and $\theta =\theta _{0}$ for some $\theta _{0}$
produces the basic ingredient for Monte Carlo tests with $H_{0}:\theta
_{T}=\theta _{0}$ where $\theta _{T}$ stands for the true value of the
parameter of interest. Changing $\theta _{0}$ for other values of the
parameter of interest produces power curves as functions of the level of the
test. This is the well known principle of Monte Carlo tests, and such is the
goal of the present section. We consider a steep but not necessarily regular
exponential family exponential family $\mathcal{P}:=\{P_{\theta ,\eta
},(\theta ,\eta )\in \mathcal{N}\}$ defined on $\mathbb{R}$ with canonical
parametrization $(\theta ,\eta )$ and minimal sufficient statistics $\left(
t,u\right) $ defined on $\mathbb{R}$ through the density

\begin{align}  \label{modele_exp_deux_param}
p_{\theta,\eta}(x):=\frac{dP_{\theta,\eta}\left( x\right) }{dx}=\exp\left[
\theta t(x)+\eta u(x)-K(\theta,\eta)\right] h(x).
\end{align}

For notational conveniency and without loss of generality both $\theta $ and 
$\eta $ belong to $\mathbb{R}$. Also the model can be defined on $\mathbb{R}%
^{d}$, $d>1$, at the cost of similar but more envolved tools. The natural
parameter space is $\mathcal{N}$ (which is a convex set in $\mathbb{R}^{2}$)
defined as the effective domain of 
\begin{equation}
k(\theta ,\eta ):=\exp \left[ K(\theta ,\eta )\right] =\int \exp \left[
\theta t(x)+\eta u(x)\right] h(x)dx.  \label{modexp}
\end{equation}

Let $X_{1}^{n}:=\left(X_{1},...,X_{n}\right) $ be $n$ i.i.d. replications of
a general random variable $\mathbf{X}$ with density (\ref%
{modele_exp_deux_param}). Denote 
\begin{equation}
T_{1,n}:=\sum_{i=1}^{n}t(X_{i})\mathit{\ \ }\text{and}\mathit{\ \ \ }%
U_{1,n}:=\sum_{i=1}^{n}u(X_{i}).  \label{TetU}
\end{equation}

Basu (1977) \cite{Basu1977} discusses ten different ways for eliminating the
nuisance parameters, among which conditioning on sufficient statistics and
consider UMPU tests pertaining to the parameter of interest. In most cases,
the density of $T_{1,n}$ given $U_{1,n}=u_{1,n}$ is unknown. Two main ways
have been developped to deal with this issue: approximating this conditional
density of a statistics or simulating samples from the conditional density.
We compose these two approaches in the present paper.

The classical technique is to approximate this conditional density using
some expansion. Then integration produces critical values. For example,
Pedersen (1979) \cite{Pedersen1979} states the mixed Edgeworth-saddlepoint
approximation, or the single saddlepoint approximation. However, the main
issue of this technique is that the approximated density still depends on
the nuisance parameter. In order to obtain the expansion, some suitable
values for the parameter of interest and for the nuisance parameter have to
be chosen. In the method developped here, as seen before, the conditional
approximated density inherits of the invariance with respect to the nuisance
parameter when conditioning on a sufficient statistics pertaining to this
parameter.

Rephrasing the notation of Section 2 in the present setting it holds that
the MLE $\left(\theta_{ML},\eta_{ML}\right)$ satisfies

\begin{equation*}
\left. \frac{\partial K\left( \theta,\eta\right) }{\partial\eta}\right\vert
_{\theta_{ML},\eta_{ML}}=u_{1,n}/n
\end{equation*}
and therefore $u_{1,n}/n$ converges to $\left( \frac{\partial K\left(
\theta_{T},\eta\right) }{\partial\eta}\right) ^{-1}\left( \eta_{T}\right) .$

For notational clearness denote $\mu$ the expectation of $u\left( \mathbf{X}%
_{1}\right) $ and $\sigma^{2}$ its variance under $\left(
\theta_{T},\eta_{T}\right) $ , hence 
\begin{equation*}
\mu:=\mu_{\left( \theta_{T},\eta_{T}\right) }:=\partial
K(\theta_{T},\eta_{T})/\partial\eta\text{ \ \ \ \ }\sigma^{2}:=\sigma_{%
\left( \theta _{T},\eta_{T}\right)
}^{2}:=\partial^{2}K(\theta_{T},\eta_{T})/\partial \eta^{2}\text{ }
\end{equation*}

Assume at present $\theta_{T}$ and $\eta_{T}$ known. It holds 
\begin{align*}
\phi(r) & :=E_{\left( \theta_{T},\eta_{T}\right) }\exp[ru\left( \mathbf{X}%
\right)] =\exp\left[ K(\theta_{T},\eta_{T}+r)-K(\theta_{T},\eta _{T})\right]
\end{align*}
and 
\begin{align*}
m(r) & =\mu_{\left( \theta_{T},\eta_{T}+r\right) } \\
s^{2}(r) & =\sigma_{\left( \theta_{T},\eta_{T}+r\right) }^{2} \\
\mu_{3}(r) & =\partial^{3}K(\theta_{T},\eta_{T}+r)/\partial\eta^{3}\text{ .}
\end{align*}

Further 
\begin{equation}
\pi _{u,\theta _{T},\eta _{T}}^{\alpha }(x):=\frac{\exp ru(x)}{\phi (r)}%
p_{\left( \theta _{T},\eta _{T}\right) }\left( x\right) =p_{\left( \theta
_{T},\eta _{T}+r\right) }\left( x\right)   \label{tiltee}
\end{equation}%
for any given $\alpha $ in the range of $P_{\theta _{T},\eta _{T}}.$ In the
above formula (\ref{tiltee}) the parameter $r$ denotes the only solution of
the equation 
\begin{equation*}
m(r)=\alpha .
\end{equation*}%
For large $k$ depending on $n,$ using Monte Carlo tests based on runs of
length $k$ instead of $n$ does not affect the accuracy of the results.

\subsection{Application of conditional sampling to MC tests}

Consider a test defined through $H_{0}:\theta _{T}=\theta _{0}$ versus $%
H_{1}:\theta _{T}\neq \theta _{0}$ Monte Carlo (MC) tests aim at obtaining $%
p-$values through simulation. where the distribution of the desired test
statistics under $H_{0}$ is either unknown or very cumbersome to obtain; a
comprehensive reference is J\"{o}ckel(1986), \cite{Jockel1986}.

Recall  the principle of thoses tests: denote $t$ the observed value of the
studied statistic\ based on the dataset and let $t_{2},..,t_{L}$ the values
of the resulting test statistics obtained through the simulation of $L-1$
samples $\mathbf{X}_{1}^{n}$ under $H_{0}.$ If $t$ is the $M$th largest
value of the sample $(t,t_{2},...,t_{L})$, $H_{0}$ will be rejected at the $%
\alpha =M/L$ signifiance level, since the rank of $t$ is uniformly
distributed on the integer $2,...,L$ when $H_{0}$ holds. Calculation of
power functions can be handled similarly. The above approximation of the
conditional density $p_{u_{1,n}}\left( x_{1}^{k}\right) $ involves the
unknown parameters $\theta _{T}$ and $\eta _{T}$ in all the simulation
steps. This problem is solved when simulating under $H_{0}:\theta
_{T}=\theta _{0}$ setting $\theta _{0}$ in place of $\theta _{T}$ and $\hat{%
\eta}_{\theta _{0}}$ in place of $\eta _{T}$, where $\hat{\eta}_{\theta _{0}}
$ is the MLE of $\eta _{T}$ in the one parameter family $p_{\theta _{0},\eta
}$ defined through (\ref{modele_exp_deux_param}). This choice follows the
commonly used one, as advocated for instance in \cite{Pedersen1979} and \cite%
{PaceSalvan1992}. Innumerous simulation studies support this choice in
various contexts.

Consider the problem of testing the null hypothesis $H_{0}:\theta
_{T}=\theta _{0}$ against the alternative $H_{1}:\theta _{T}>\theta _{0}$ in
model (\ref{modexp}) where $\eta $ is the nuisance parameter.

When $p_{u_{1,n},\theta_{0}}$ is known, the classical conditional test $%
H_{0}:\theta_{T}=\theta _{0}$ versus $H_{1}:\theta _{T}>\theta _{0}$ with
level $\alpha$ is UMPU.

Substituting $p_{\theta_{0}}\left(\mathbf{X}_{1}^{n}=x_{1}^{n}\left\vert 
\mathbf{U}_{1,n}=u_{1,n}\right.\right)$ by $g_{u_{1,n},\theta_{0}}\left(
x_{1}^{k}\right) $ defined in (\ref{g_s}), i.e. substituting the test
statistics $T_{1}^{n}$ by $T_{1}^{k}$ and $p_{\theta_{0}}\left(\mathbf{X}%
_{1}^{k}=x_{1}^{k}\left\vert \mathbf{U}_{1,n}=u_{1,n}\right.\right)$ by $%
g_{u_{1,n},\theta_{0}}\left(x_{1}^{k}\right) $ i.e. changing the model for a
proxy while keeping the same parameter of interest $\theta $ yields the
conditional test with level $\alpha $

\begin{equation*}
\psi _{\alpha }(x_{1}^{k}):=\left\{ 
\begin{array}{r}
1\text{\ \ if\ \ }T_{1,k}>t_{\alpha } \\ 
\gamma \text{\ \ if\ \ }T_{1,k}=t_{\alpha } \\ 
0\text{\ \ if\ \ }T_{1,k}<t_{\alpha }%
\end{array}%
\right.
\end{equation*}%
and 
\begin{equation*}
E_{G_{u_{1,n}}}[\psi _{\alpha }(X_{1}^{k})]=\alpha
\end{equation*}%
i.e. $\alpha :=\int \mathds{1}_{t_{1,k}>t_{\alpha}}g_{u_{1,n}}\left(
x_{1}^{k}\right) dx_{1}...dx_{k}.$ Its power under a simple hypothesis $%
\theta _{T}=\theta $ is defined through 
\begin{equation*}
\beta _{\psi _{\alpha }}(\theta |u_{n})=E_{\theta }[\psi
_{n}(T_{1,n},U_{1,n})|U_{1,n}=u_{1,n}].
\end{equation*}

Recall that the parametric bootstrap produces samples from a parametric
model which is fitted to the data, often through maximum likelihood. In the
present setting, the parameter $\theta $ is set to $\theta _{0}$ and the
nuisance parameter $\eta $ is replaced by its estimator $\widehat{\eta }%
_{\theta _{0}}$ which is the MLE of $\eta $ when the parameter $\theta $ is
fixed at the value $\theta _{0}$ defining $H_{0}.$ Comparing their exact
conditional MC tests with parametric bootstrap ones for Gamma distributions,
Lockhart et al(2007)\cite{LOckhart2005} conclude that no significant
difference can be notices in terms of level or in terms of power. We proceed
in the same vein, comparing conditional sampling MC tests with the
parametric bootstrap ones, obtaining again similar results when the nuisance
parameter is estimated accurately. However the results are somehow different
when the nuisance parameter cannot be estimated accurately, which may occur
in various cases.

In practice since the chosen conditioning statistics is quasi sufficient for
the nuisance parameter, we plug any value for this parameter in the
definition of $g_{u_{1,n}}.$ This is what has been performed in all examples
below.

\subsection{Unimodal Likelihood: testing the coefficients of a Gamma
distribution}

Let $\mathbf{X}_{1}^{n}$ be an i.i.d. sample of random variables with Gamma
distribution $\Gamma\left( a_{T},b_{T}\right) $ where $a_{T}$ is the shape
coefficient and $b_{T}$ is the scale coefficient. As $a$ and $b$ vary this
distribution is a two parameter exponential family. The statistics $T_{1,n}:=%
\mathbf{X}_{1}+...+\mathbf{X}_{n}$ is sufficient for the parameter $a$ and $%
U_{1,n}:=\log\mathbf{X}_{1}+...+\log\mathbf{X}_{n}$ is sufficient for $b.$

\paragraph*{MC conditional test with $H_{0}:a_{T}=a_{0}$}

Denote $u_{1,n}=\sum_{i=1}^{n}X_{i}$ and $\hat{b}_{a_{0}}$ the MLE of $b.$
Calculate for $l\in \{2,L\}$ 
\begin{equation*}
t_{l}:=\sum_{i=0}^{k}\log \left( Y_{i}(l)\right) .
\end{equation*}%
where the $Y_{i}^{\prime }$ are a sample from $g_{u_{1,n}}^{a_{0},\hat{b}%
_{a_{0}}}.$

Consider the corresponding parametric bootstrap procedure for the same test,
namely simulate $Z_{i}(l)$, $2\leq l\leq L$ and $0\leq i\leq k$ with
distribution $\Gamma \left( a_{0},\hat{b}_{a_{0}}\right) ;$ denote 
\begin{equation*}
s_{l}:=\sum_{i=0}^{k}\log \left( Z_{i}(l)\right) .
\end{equation*}

In this example simulation shows that for any $\alpha $ the $M$th largest
value of the sample $\left( t,t_{2},...,t_{L}\right) $ is very close to the
corresponding empirical $M/L$-quantile of $s_{l}$'s. Hence Monte Carlo tests
through parametric bootstrap and conditional compete equally. Also in terms
of power, irrespectively in terms of $\alpha $ and in terms of alternatives
(close to $H0$), the two methods seem to be equivalent.

\paragraph*{MC conditional test with $H_{0}:b_{T}=b_{0}$}

Denote $u_{1,n}=\sum_{i=1}^{n}\log \left( X_{i}\right) $ and $\hat{a}%
_{b_{0}} $ the MLE of $a.$ Calculate for $l\in \{2,L\}$ 
\begin{equation*}
t_{l}:=\sum_{i=0}^{k}Y_{i}(l)
\end{equation*}%
where the $Y_{i}^{\prime }$ are a sample from $g_{u_{1,n}}^{b_{0},\hat{a}%
_{b_{0}}}$ and, as above define accordingly 
\begin{equation*}
s_{l}:=\sum_{i=0}^{k}\log \left( Z_{i}(l)\right)
\end{equation*}%
where the $Z_{i}(l)$'s are simulated under $\Gamma \left( \hat{a}%
_{b_{0}},b_{0}\right) .$

As above, parametric bootstrap and conditional sampling yield equivalent
Monte Carlo tests in terms of power function under alternatives close to $H0$%
.

In the two cases studied above \ the value of $k$ has been obtained through
the rule exposed in section 3.2 of Broniatowski and Caron (2011) \cite%
{BroniaCaronAAP2011}.

\subsubsection{Bimodal likelihood: testing the mean of a normal distribution
in dimension 2}

In contrast with the above mentionned examples, the following case study
shows that estimation through the unconditional likelihood may fail to
provide consistent estimators when the likelihood surface has multiple
critical points. This in turn yields parametric bootstrap Monte Carlo tests
with inacceptable power functions.\ 

Sundberg(2009)\cite{Sundberg2009} proposes four examples that allow
likelihood multimodality. Two of them can also be found in \cite{Efron1975}
and \cite{Efron1978}, and in \cite{BardnoffCox1994}, Ch 2. We consider the
"Normal parabola" model which is a curved (2, 1) family (see Example 2.35 in 
\cite{BardnoffCox1994}, Ch 2 ). Two independent Gaussian variates have
unknown means and known variances; their means are related by a parabolic
relationship.

Let $X$ et $Y$ be two independent gaussian r.v.'s with same variance $\sigma
_{T}^{2}$ with expectation $\psi _{T}$ and $\psi _{T}^{2}$. In the present
example $\sigma _{T}^{2}=1$ and $\psi _{T}=2.$

Let $\left( X_{i},Y_{i}\right) $ , $1\leq i\leq n$ be an i.i.d. sample with
the above distribution.

The parameter of interest is $\sigma^{2}$ whislt the nuisance parameters is $%
\psi.$ Derivation of the likelihood function of the observed sample with
respect to $\psi$ yields the following equation 
\begin{equation*}
\left( U_{1,n}-\psi\right) +2\psi\left( V_{1,n}-\psi^{2}\right) =0
\end{equation*}
with $U_{1,n}:=X_{1}+...+X_{n}$ and $V_{1,n}:=Y_{1}+...+Y_{n}.\ $The
following table shows that the likelihood function is bimodal in $\psi.$

\begin{figure}[h!t]
\centering \includegraphics[width=5cm]{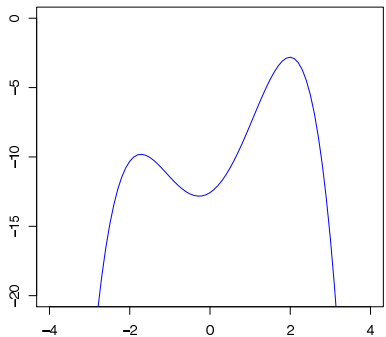}
\caption{Bimodal likelihood in $\protect\psi.$}
\end{figure}

Estimation of the nuisance parameter $\psi $ is performed through the
standard Newton Raphson method. The Newton-Raphson optimizer of the
likelihood function converges to the true value when the initial value is
larger than 1and fails to converge to $\psi _{T}=2$ otherwise. Hencefore the
parametric bootstrap estimation of the likelihood function of the sample
based on this preliminary estimate of the nuisance parameter may lead to
erroneous estimates of the parameter of interest. Indeed \ according to the
initial value we obtained estimators of $\psi _{T}$ close to $2$ or to $-2.$
When the estimator of the nuisance parameter is close to its true value $2$
then parametric bootstrap yields Monte Carlo tests with power close to $1$
for any $\alpha $ and any alternative close to $H0.$ At the contrary when
this estimate is close to the second maximizer of the likelihood (i.e. close
to $-2$) then the resulting Monte Carlo test based on parametric bootstrap
has power close to $0$ irrespectively of the value of $\alpha $ and of the
alternative, when close to $H0.$ In contrast with these results, Monte Carlo
tests based on conditional sampling provide powers close to $1$ for any $%
\alpha $; we have considered alternatives close to $H0$ $.$ This result is
of course a consequence of quasi sufficiency of the statistics $\left(
U_{1,n},V_{1,n}\right) $ for the parameter $\psi $ of the distribution of
the sample $\left( X_{i},Y_{i}\right) _{i=1,...,n}$; see next paragraph for
a discussion of this point.

\subsection{Estimation through conditional likelihood}

\label{sec:cond_like}

Considering model (\ref{modexp}) we intend to perform an estimation of $%
\theta_{T}$ irrespectively upon the value of $\eta _{T}$. Denote $\widehat{%
\eta}_{\theta }$ the MLE of $\eta _{T}$ when $\theta $ holds; the model $%
p_{\theta ,\widehat{\eta }_{\theta }}(x)$ is a one parameter model which is
fitted to the data for any peculiar choice of $\theta.$ The classic
unconditional likelihood provides consistent estimators of $\theta _{T}$ in
many cases. However, this method strongly relies on the constistency
properties of $\widehat{\eta }_{\theta }$ at any given $\theta .$

For fixed parameter value $\theta$ of the parameter of interest, Theorem 1
means that the likelihood of the subsample $\mathbf{X}_{1}^{k}$ with unknown
distribution with parameter $\left( \theta _{T},\eta _{T}\right) $ under the
distribution with any parameter $\theta $ given the value of the sufficient
statistics $U_{1,n}$ is approximated by $g_{u_{1,n}}\left( \mathbf{X}%
_{1}^{k}\right) $ when $\mathbf{X}_{1}^{k}$ is either generated under the
conditional density or under $g_{u_{1,n}}$ $\ $with parameter $\eta =\eta
_{T}.$ Substituting $\eta _{T}$ by its estimator should yield maximal value
of the approximate likelihood when $\left(\theta _{T},\eta _{T}\right)$
holds, since $\widehat{\eta}_{\theta}$ approaches $\eta _{T}$ when $\theta
=\theta _{T}.$ In particular, this holds when $\mathbf{X}_{1}^{k}$ is
generated under $\theta _{T},\eta _{T}$ which holds on the observed sample.
This yields to an algorithm to estimate $\theta _{T}.$ For any $\theta $
calculate $\widehat{\eta }_{\theta }$ $.$ Evaluate $g_{u_{1,n}}\left( 
\mathbf{X}_{1}^{k}\right) $ and optimize in $\theta.$

In most cases, as the normal, gamma or inverse-gaussian, both estimation
through the unconditional likelihood and estimation through conditional
likelihood based on the proxy $g_{u_{1,n}}$ give a consistent estimator.

We consider the example of the Bimodal likelihood from the above subsection,
inheriting of the notation and explore the behaviour of the proxy of the
conditional likelihood of the sample $\left( X_{i},Y_{i}\right) $ , $1\leq
i\leq n$ when conditioning on $U_{1,n}$ and $V_{1,n}$ , as a function of $%
\sigma ^{2}.$ The likelihood writes 
\begin{align*}
& L\left( \left. \sigma ^{2}\right\vert \left( X_{i},Y_{i}\right) ,1\leq
i\leq n,U_{1,n},V_{1,n}\right) \\
& =P_{\mathbf{X}_{1}^{n}}\left( \left. X_{1}^{n}\right\vert U_{1,n},\sigma
^{2}\right) P_{\mathbf{Y}_{1}^{n}}\left( \left. Y_{1}^{n}\right\vert
V_{1,n},\sigma ^{2}\right)
\end{align*}%
where we have used the independence of the r.v.'s $X_{i}$'s and $Y_{i}$'s.

\begin{figure}[tbp]
\centering \includegraphics*[ scale=0.4]{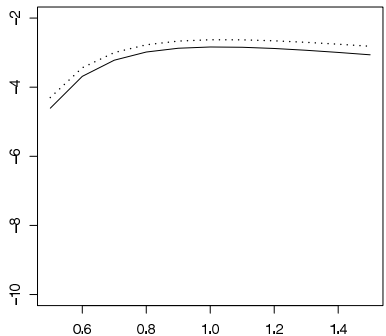}
\caption{Proxy of the conditional likelihood (solid line) along with the
empirical likelihood (dotted line) as function of $\protect\sigma^{2}$ for $%
n=100$ and $k=99$ in the case where a good initial point in Newton-Raphson
procedure is chosen.}
\label{table:NR_good_start}
\end{figure}

Applying Theorem 1 to the above expression it appears that $\psi $ cancels
in the resulting density $g_{u_{1,n}}$ and $g_{v_{1,n}}.$ This proves that
the proxy of the conditional likelihood provides consistent estimation of $%
\sigma _{T}^{2}$ as shown on Figures \ref{table:NR_good_start} and \ref%
{table:NR_bad_start} (see the solid lines).

\begin{figure}[tbp]
\centering \includegraphics*[ scale=0.4]{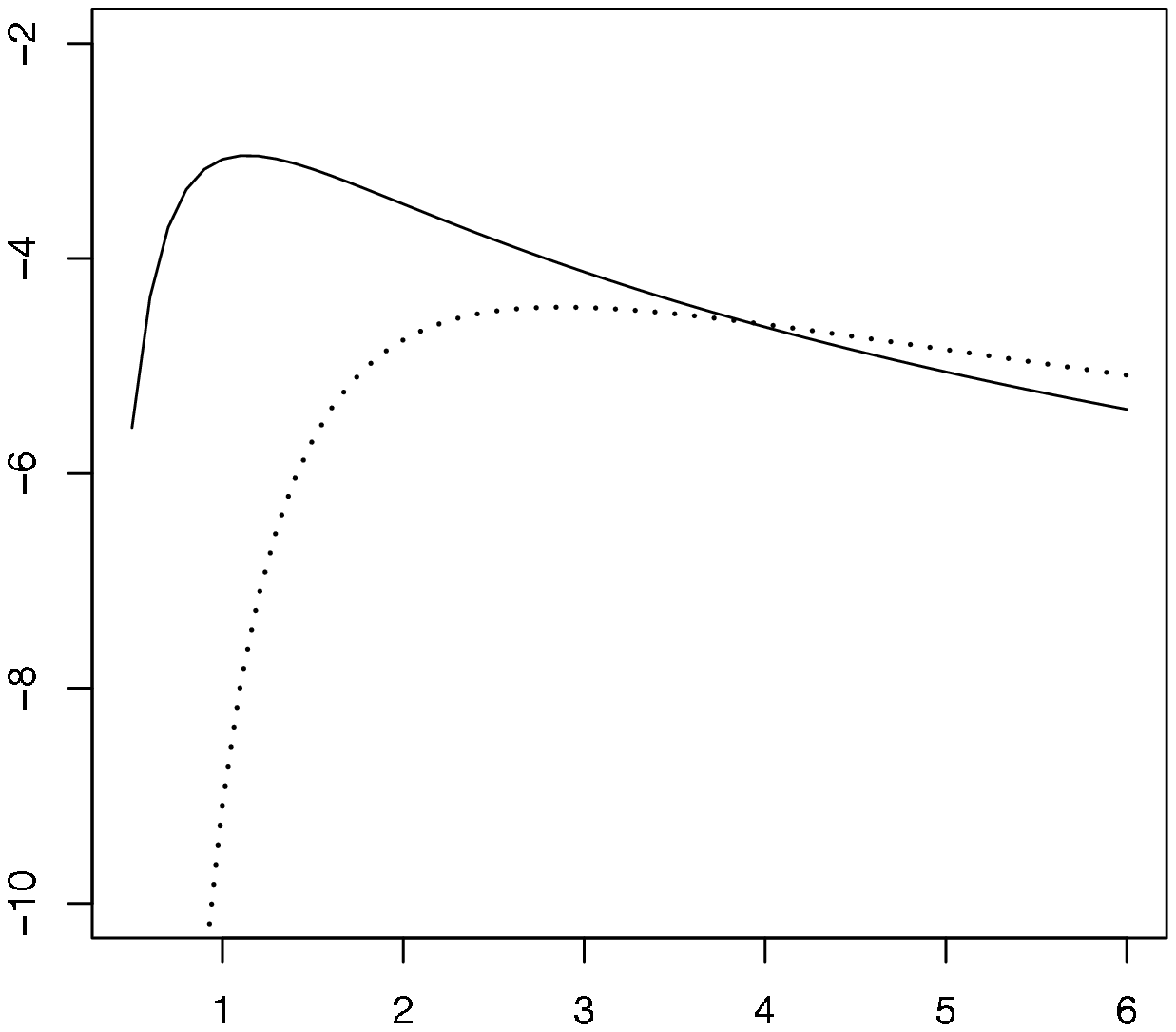}
\caption{Proxy of the conditional likelihood (solid line) along with the
empirical likelihood (dotted line) as function of $\protect\sigma^{2}$ for $%
n=100$ and $k=99$ in the case where a bad initial point in Newton-Raphson
procedure is chosen.}
\label{table:NR_bad_start}
\end{figure}

On Figure \ref{table:NR_good_start}, the dot line is the empirical
likelihood function with consistent estimator of the nuisance parameter; the
resulting maximizer in the variable $\sigma ^{2}$ is close to $\sigma
_{T}^{2}=1.$ At the opposite in Figure \ref{table:NR_bad_start} an
inconsistent preliminary estimator of $\psi _{T}$ obtained through a bad
tuning of the initial point in the Newton-Raphson procedure leads to
unconsistency in the estimation of $\sigma _{T}^{2}$, the resulting
likelihood function being unbounded.

\end{document}